%Paper: cond-mat/9507033
%From: John.Mac-Cabe@enslapp.ens-lyon.fr (John MCCABE)
%Date: Wed, 12 Jul 95 12:45:57 +0200

\baselineskip=15pt
\magnification=1200

\centerline{\bf CRITICAL CORRELATIONS OF THE 2-DIMENSIONAL, 3-STATE POTTS
MODEL}

\vskip 1cm

\centerline{John McCabe}

\vskip .2cm

\centerline{\it Laboratoire de Physique Th\'eorique ENSLAPP\footnote{$^1$}{URA
14-36 du CNRS, associ\'ee \`a l'Ecole Normale Sup\'erieure de Lyon et \`a
l'Universit\'e de Savoie.}}
\centerline{\it ENS Lyon, 46 all\'ee d'Italie, 69364 Lyon Cedex 07, France}

\vskip .4cm

\centerline{Tomasz Wydro}

\vskip .2cm

\centerline{\it LPLI-IPC}
\centerline{\it Universit\'e de Metz}
\centerline{\it 1 bd Arago, 57070 Metz, France}

\vskip 1cm

\centerline{\bf ABSTRACT}

\vskip .4cm

We exploit the identification between the critical theory of the 3-state Potts
model and the $D_5$ conformal model. This allows us to determine all 3-point
correlations involving the fields associated with the Potts order parameter
and the energy density. The calculation uses screened Coulomb gas
correlations. The $\bf Z_3$ symmetry of the 3-state Potts model is manifestly
preserved.

\vskip 3cm

\rightline{ENSLAPP-L-526/95}
\rightline{June 1995}

\vfill\eject

Sometime ago Baxter showed that the 3- and 4-state Potts models, in 2
dimensions, have second-order phase transitions [1]. A futher identification
with the  hard hexagon model gave the critical exponents of the 3-state model
[2].  Knowledge of these exponents motivated Dotsenko to try to calculate the
correlations by exploiting the identification of 2-dimensional, critical models
and conformal field theories [3,4].

Dotsenko's study of the Potts correlation functions was made before the
complete classification of minimal conformal field theories [5]. Thus, he
attempted to construct the 3-state Potts correlations from the then well-known
$A_5$ conformal theory. The $A_5$ model has one deficiency as a theory for the
Potts model. It has only one real field with exponent $\eta=4/15$ whereas the
Potts model has two fields with this value of $\eta$, the real and imaginery
parts of the complex order parameter $\sigma (\vec x)$. Thus, the $A_5$ model
is unable to represent all the correlation functions of the Potts model. To be
more precise, the $A_5$ model cannot be used to calculate the Potts correlation
$<\sigma(\vec x_1)\sigma(\vec x_2)\sigma(\vec x_3)_+>$. This correlation is
intimately related to the complex nature of $\sigma (\vec x)$ and its
transformations under $\bf Z_3$. The $D$-series conformal theories, unlike the
$A$-series, allow complex discrete symmetries.

This article extends previous work on the $D_5$ conformal field theory [6].
There it was shown that the operator algebra of $D_5$ could be constructed to
respect the discrete symmetry $\bf Z_2$. Here, we show that there is a second
solution which has the discrete symmetry $\bf Z_3$. The two solutions are, of
course, inequivalent. \footnote{$^1$}{It appears that there are two different
$D_5$ models with either the discrete symmetry group $\bf Z_2$ or $\bf Z_3$.}
The new solution gives the critical correlations of the 3-state Potts model. We
obtain the nonzero structure constants involving any combinations of the Potts
order parameter and the energy density fields. The results, which are given in
an exact form in Table 2, are that $C_{\epsilon,\epsilon,\epsilon}=0$,
$C_{\sigma^*,\sigma,\epsilon}\simeq .546$, and $C_{\sigma,\sigma,\sigma}\simeq
1.092$.

\vskip .2cm

\leftline{\bf 1. Introduction}

\vskip .2cm

We summarize some facts about the Potts model which enable the identification
of its critical point with the $D_5$ conformal theory.  The lattice Hamiltonian
of the {\it q}-state Potts model is [7]:

$$H=J\sum_{\vec x, \vec {\alpha}}\{\sigma (\vec x)\sigma^*(\vec x +\vec
{\alpha})\  + \ \sigma^* (\vec x)\sigma(\vec x +\vec {\alpha})\}\eqno{(1)}$$
The variable $\sigma(\vec x)$ is the complex order parameter. The order
parameter takes its values among the {\it q}th roots of unity.
For the 3-state model, $\sigma(\vec x)\epsilon\{1, \exp{\left({\pm i2\pi\over
3}\right)}\}$. The vectors $\alpha_i$ generate the model's lattice. The Potts
Hamiltonian is invariant under the discrete symmetry group $\bf Z_q$ which is
realized by the transformation $\sigma (\vec x) \buildrel {\bf Z_q} \over
\longrightarrow \exp{\left({i2\pi\over q}\right)}\sigma(\vec x)$. The existence
of this symmetry at the critical point is the reason that the 3-state Potts
model is described by the $D_5$ conformal theory and not the $A_5$ conformal
theory.

To make the identification between the conformal model and the critical Potts
model precise, it is necessary to know: the Potts model's central charge at
$T_c$, the conformal dimensions of scaling fields like $\sigma (\vec x)$ and
the energy density $\epsilon (\vec x)$, and tranformation properties of scaling
fields under the discrete symmetries of $\bf Z_q$.

The central charge can be calculated with arguments based on the
transformations of energy momentum tensor under the conformal group. The
conformal mapping of the cylinder onto the plane allows us to calculate the
ground state energy of a conformal model defined on an infinitely long cylinder
of radius R. It has the form  $E_o={-c\over 12R}$ where $c$ is the central
charge of the conformal algebra [8]. Thus, the central charge of the conformal
theory is directly calculable from the ground state energy of the statistical
system at the critical temperature $T_c$. The ground state energy of the Potts
model on a cylinder has been calculated [9]. Its value leads to the conclusion
that $c=4/5$ for the 3-state Potts model.

As stated above, the exponents defining the behavior near $T_c$, for the
scaling field $\sigma (\vec x)$ and $F(T)$, are know by the relationship
between the Potts model and the hard hexagon model [2]. For the 3-state Potts
model the identification fixes the exponents: $\alpha$, $\beta_{\sigma}$ and
$\delta$.
$${F(T)/kT}\sim (T-T_c)^{2-\alpha}, <\sigma>\sim (T-T_c)^{\beta_{\sigma}}\ {\rm
with}\ \ \alpha=1/3,\ \beta_{\sigma}=1/9, \ \delta=14\eqno{(2)}$$
Furthermore, the relation between the energy density and the Helmholtz function
$F(T)$, allow us to extract the exponent $\beta_{\epsilon}$ of the second
scaling field $\epsilon (\vec x)$, the energy density.

$$ <\epsilon>=-kT^2\partial_T\left(F(T)/kT\right)
\  {\rm and}\ <\epsilon>\sim (T-T_c)^{\beta_{\epsilon}}\ \ \eqno{(3)}$$
Thus, $\beta_{\epsilon}=1-\alpha$, or $\beta_{\epsilon}=2/3$ for the 3-state
Potts model.

We will need the exponents $\eta$ and {\it s} (spin) which define the 2-point
correlation function at the critical temperature.\footnote{$^2$}{The complex
coordinates $z\equiv x+iy$ will be more convenient for the remainder of this
article.}
$$<\Phi_i(z, \bar z)\Phi^*_j (0,0)_+>={\displaystyle \delta_{ij}(\bar z /
z)^{s_i}\over (z\bar z)^{{\eta_i}/2}}={\delta_{ij}\over z^{2\Delta_i}\bar
z^{2\bar\Delta_i}} \ \ \eqno{(4)}$$
For the two spinless fields $\sigma (z,\bar z)$ and $\epsilon (z,\bar z)$,
standard scaling relations allow us to calculate the relevant $\eta_i$'s [8].
$$\alpha+2\beta_i+\gamma_i=2,\ \ 2\nu=2-\alpha,\ \ {\rm and}\ \
(2-\eta_i)\nu=\gamma_i \ \ \eqno{(5)}$$
They imply that $\eta_{\sigma}=4/15$ and $\eta_{\epsilon}=8/5$ in the 3-state
Potts model. The exponents $\eta$ and $s$ define the two conformal dimensions
of the primary fields associated with the scaling fields: $\Delta={\eta +
2s\over 4}$ and $\bar \Delta={\eta - 2s\over 4}$.
Thus, we see that the conformal field theory associated with the critical
3-state Potts model must have, at least, the following primary fields
($\Delta,\bar \Delta$): 2 real primary fields (1/15,1/15) associated with
$\sigma_{real}$ and $\sigma_{imaginery}$, and 1 real primary field (2/5,2/5)
associated with $\epsilon$.

The form of the spectrum of primary fields of the conformal theory associated
to Potts follows from three constraints. First, the unitarity of the 3-state
Potts model's transfer matrix fixes the spectrum of conformal dimensions
$\Delta$ and $\bar\Delta$ to be those of the degenerate spectrum of Kac [10],
i.e. $\Delta_{r,s}={(r(m+1)-sm)^2-1\over 4m(m+1)}$ where $c=1-{6\over m(m+1)}$.
Second, finite size scaling arguments require that the thermal partition with
toridal boundary conditions be modular invariant [11]. Since  $c=4/5$ and $m=5$
for the 3-state Potts model [9], this restricts the spectrum of dimensions
($\Delta_{rs}, \bar\Delta_{r^\prime s^\prime}$) to those of either the $A_5$ or
the $D_5$ minimal models [5]. The final constraint that chooses between these
two models is the $\bf Z_3$ symmetry of the Potts model. Only the $D_m$ models
have inequivalent primary fields with the same conformal dimensions, i.e. for
($\Delta_{r,{m+1\over 2}}, \bar\Delta_{r,{m+1\over 2}}$) with $1\le r<s\le m$.
The field

$\sigma (z,\bar z)$, which forms a complex representation of $\bf Z_3$, must be
represented by two inequivalent primary fields of conformal dimension($1/15,
1/15$). Thus, the 3-state Potts model is desribed by the $D_5$ conformal model,
and $\sigma (z,\bar z)$ is represented by two real ($\Delta_{2,3},
\bar\Delta_{2,3}$) primary fields. This conclusion will generalize to the
statement that all critical points having complex representations of discrete
symmetries (and $c<1$) are described by $D$-series conformal theories.

The complete spectrum of primary fields for the $D_5$ model is [5,6]:
\vskip .4cm
\centerline{TABLE 1: Operator Content of $D_5$ Conformal Model}
\vbox{\offinterlineskip\hrule
\halign{&\vrule#&\strut\quad\hfil#\hfil\quad\cr
height2pt&\omit&&\omit&&\omit&&\omit&&\omit&&\omit&\cr
&($r,s|r^\prime,s^\prime$)&&$\Delta$&&$\bar\Delta$&&Multiplicity&&{$\bf Z_3$}
singlet? && Critical Field&\cr
height2pt&\omit&&\omit&&\omit&&\omit&&\omit&&\omit&\cr
\noalign{\hrule}
height2pt&\omit&&\omit&&\omit&&\omit&&\omit&&\omit&\cr
&($11|11$)&&0&&0&&1&&yes&&\omit&\cr
&($13|13$)&&2/3&&2/3&&2&&\bf no&&$\Omega$&\cr
&($15|15$)&&3&&3&&1&&yes&&\omit&\cr
&($15|11$)&&3&&0&&1&&yes&&\omit&\cr
&($11|15$)&&0&&3&&1&&yes&&\omit&\cr
&($21|21$)&&2/5&&2/5&&1&&yes&&$\epsilon$&\cr
&($23|23$)&&1/15&&1/15&&2&&\bf no&&{$\sigma$}&\cr
&($25|25$)&&7/5&&7/5&&1&&yes&&\omit&\cr
&($21|25$)&&2/5&&7/5&&1&&yes&&\omit&\cr
&($25|21$)&&7/5&&2/5&&1&&yes&&\omit&\cr
height2pt&\omit&&\omit&&\omit&&\omit&&\omit&&\omit&\cr}
\hrule}
The Potts fields $\sigma$ and $\epsilon$ and a third field, $\Omega$, are
indicated in the list of primary fields of $D_5$. From the operator product
expansion, one can show that both $\Omega$ and $\sigma$ transform under the
$\bf Z_3$ symmetry group of the Potts model. We shall always use the complex
combinations of the two independent real primary fields $\Phi^i_{(13|13)}$ and
$\Phi^i_{(23|23)}$ defined by: $\sigma(z,\bar z)\equiv\Phi^1_{(23|23)}(z,\bar
z) +{\bf i}\Phi^2_{(23|23)}(z,\bar z)$ and $\Omega(z,\bar
z)\equiv\Phi^1_{(13|13)}(z,\bar z) +{\bf i}\Phi^2_{(13|13)}(z,\bar z)$ . On
these fields, the discrete $\bf Z_3$ symmetry group of the 3-state Potts model
is realized by:
$$ \sigma(z,\bar z){\buildrel{\bf Z_3}\over\longrightarrow}
e^{i2\pi\over3}\sigma(z,\bar z)\ \ {\rm and}\ \
\Omega(z,\bar z){\buildrel{\bf Z_3}\over\longrightarrow}
e^{i2\pi\over3}\Omega(z,\bar z)
\ \ .\eqno{(6)}$$
The implementation of the $\bf Z_3$ symmetry will be crucial in solving the
bootstrap equations for 4-point functions. This completes the identification of
the correct conformal theory for the critical Potts model.

\leftline{\bf 2. Fusion Rules and Operator Product Expansions}

Having determined how the $D_5$ conformal model represents the critical fields
of the 3-state Potts model, we turn to the calculation of correlation
functions. Our calculation of structure constants will use the methods of
Dotsenko and Fateev [12]. The extension to non-diagonal models is discussed in
Ref. [6]. We will rely on these articles heavily and only give the details
necessary to clarify the discussion.

In conformal theories, the correlation of 3 primary fields takes the following
form [13].
$$<\Phi_{\Delta_a,\bar\Delta_a}(1)\Phi_{\Delta_b,\bar\Delta_b}(2)\Phi_{\Delta_c,\bar\Delta_c}(3)_+>={C_{a,b,c}\over (z_{12})^{\Delta_1+\Delta_2-\Delta_3}
(\bar z_{12})^{\bar\Delta_1+\bar\Delta_2-\bar\Delta_3}\times cyclic \ perm.}
\eqno{(7)}$$
The constants $C_{a,b,c}$ are called the structure constants. When the 2-point
correlations are normalized as in (4), the structure constants are universal
quantities. They are the interest in this article. These constants also appear
as coefficients in the expansions of the products of primary
fields\footnote{$^3$}{The fields $\sigma^*(z,\bar z)$ and $\Omega^*(z,\bar z)$
are defined through the replacement ${\bf i}\longrightarrow {\bf -i}$ in the
definitions of $\sigma(z,\bar z)$ and $\Omega(z,\bar z)$.}
 [3], i.e. $(\Phi_a(1)\times\Phi_b(2))_+\simeq\sum_cC_{a,b,c}\Phi^*_{c}(2)/
z_{12}^{\Delta_a+\Delta_b-\Delta_c}\bar
z_{12}^{\bar\Delta_a+\bar\Delta_b-\bar\Delta_c}$.

The transformations of the structure constants under permutations of the three
fields are also easily obtained from (7). The $C_{a,b,c}$'s are invariant under
cyclic permutations of the indicies $a,b,$ and $c$. They are related by a sign
under exchanges of two indices.
$$C_{a,b,c}=(-1)^{[\Delta_a-\bar\Delta_a+\Delta_b-\bar\Delta_b+\Delta_c-\bar\Delta_c]}C_{b,a,c}\eqno{(8a)}$$
Lastly, the hermitian conjugation of conformal theories,
$\Phi^\dagger_{a}(z,\bar z)\equiv$ $z^{-2\Delta}\bar
z^{-2\bar\Delta}\Phi^*_{a}(1/z,1/\bar z)$, relates the structure constants to
their complex conjugates.
$$C^*_{a,b,c}=(-1)^{[\Delta_a-\bar\Delta_a+\Delta_b-\bar\Delta_b+\Delta_c-\bar\Delta_c]}C_{a^*,b^*,c^*}\eqno{(8b)}$$
We shall be specially interested by the result for the exchange $\sigma^*
\longleftrightarrow\sigma$ from which we see that
$C_{\sigma^*,\sigma,c}=(-1)^{\Delta_c-\bar\Delta_c}C_{\sigma,\sigma^*,c}$ and
by
$C^*_{\sigma,\sigma,c}=(-1)^{\Delta_c-\bar\Delta_c}C_{\sigma^*,\sigma^*,c^*}$.

This article employs 4-point functions to calculate the structure constants.
One must determine which 4-point correlations depend on the structure constants
desired. The Fusion Rules for primary fields having weights given by the Kac
formula ($\Delta_{r_i,s_i},\bar\Delta_{r_i^\prime,s_i^\prime}$) will help
resolve this question. The structure constants and associated operator product
coefficients are nonzero only if the Kac weights satisfy the following
conditions:
$$\eqalign{r_a+r_b+r_c=1 \ (mod \ 2),\ \ (r_a-1)+(r_b-1)\ge(r_c-1), \ \ {\rm
cyclic\ permutations}\cr
s_a+s_b+s_c=1 \ (mod \ 2),\ \ (s_a-1)+(s_b-1)\ge(s_c-1), \ \ {\rm cyclic\
permutations}\cr}\eqno{(9)}$$
The $(r^\prime_i,s^\prime_i)$ satisfy identical equations. For the primary
fields representing $\epsilon(z,\bar z)$ and $\sigma(z,\bar z)$ these
conditions limit the operator product expansions to have the following form:
$$\eqalign{&\Phi_{(21|21)}\times\Phi_{(21|21)}=\ 1\  \oplus\Phi_{(25|25)}\ ,\cr
&\Phi_{(23|23)}\times\Phi_{(21|21)}=\ \Phi_{(23|23)} \ \oplus\ \
\Phi_{(13|13)}\ ,\cr
&\Phi_{(23|23)}\times\Phi_{(23|23)}=\ 1\oplus\ \ \Phi_{(13|13)} \ \oplus\ \
\Phi_{(15|15)}\ \oplus\ \ \Phi_{(11|15)}\ \oplus\ \ \Phi_{(15|11)} \  \cr
 &\ \ \ \ \ \ \ \ \ \ \ \ \ \ \ \ \oplus\ \Phi_{(21|21)} \ \oplus\ \
\Phi_{(23|23)}\ \oplus\ \ \Phi_{(25|25)}\oplus\  \ \Phi_{(21|25)} \ \oplus
\Phi_{(25|21)}\
.\cr}\eqno{(10a)}$$
When one imposes the ${\bf Z_3}$ symmetry of the Potts model, the Fusion Rules
further simplify.
$$\eqalign{ \epsilon\times\epsilon= & 1\  \oplus\Phi_{(25|25)}\ ,\cr
\sigma\times\epsilon= &\ \sigma \ \oplus\ \Omega\ ,\cr
\sigma\times\sigma= &\ \Omega^* \ \oplus \ \sigma^*\ ,\cr
\sigma\times\sigma^*= &\ 1\oplus\ \Phi_{(15|15)}\ \oplus\ \Phi_{(11|15)}\
\oplus\ \Phi_{(15|11)}\ \oplus \epsilon\ \oplus \Phi_{(25|25)}
\ \oplus\ \Phi_{(21|25)}\ \oplus\ \Phi_{(25|21)}\ .\cr} \eqno{(10b)}$$

These expressions allow us to calculate the 4-point functions in the s- and
t-channels in a short distance expansion. It should be emphasized that the use
of (10b) in the operator product expansion of 4-point correlations is
equivalent to imposing the discrete symmetry $\bf Z_3$ on the conformal theory.
It is at this point that our analysis differs from both the $A_5$ [4] and the
$\bf Z_2$ symmetric $D_5$ models [6]. The discrete symmetry group of the
statistical model defines the conformal model.

The Fusion Rules ,(10b), are already sufficient to show that
$C_{\epsilon,\epsilon,\epsilon}=0$. To determine
$C_{\sigma,\sigma^*,\epsilon}$ and $C_{\sigma,\sigma,\sigma}$, we must study:
$<\sigma^* (1)\epsilon (2)\epsilon (3)\sigma (4)_+>$ and $<\sigma (1)\sigma
(2)\sigma^* (3)\sigma^* (4)_+>$. From standard arguments [6] and the $\bf Z_3$
symmetric Fusion Rules, (10b), their short distance expansions in the s-channel
($|z_{12}|, |z_{34}| \ll 1$) are given by:

$$<\sigma^* (1)\epsilon (2)\epsilon (3)\sigma (4)_+>\simeq {\displaystyle
\left|C_{\sigma^*,\epsilon,\sigma}\right|^2\over
|z_{12}|^{4/5}|z_{34}|^{4/5}|z_{24}|^{4/15}}\ +\ {\displaystyle
\left|C_{\sigma^*,\epsilon,\Omega}\right|^2\over
|z_{12}|^{-2/5}|z_{34}|^{-2/5}|z_{24}|^{8/3}}\ ,\eqno{(11a)}$$
and
$$<\sigma (1)\sigma (2)\sigma^* (3)\sigma^* (4)_+>\simeq\ {\displaystyle
|C_{\sigma,\sigma,\sigma}|^2\over
|z_{12}|^{2/15}|z_{34}|^{2/15}|z_{24}|^{4/15}}\ +\ {\displaystyle
|C_{\sigma,\sigma,\Omega}|^2\over
|z_{12}|^{-16/15}|z_{34}|^{-16/15}|z_{24}|^{8/3}}\ .\eqno{(11b)}$$
We will also need the operator product expansions of these correlation
functions in the t-channel ($|z_{41}|,|z_{23}|\ll 1$). Applying the Fusion
Rules of (10) again, we find that:
$$<\sigma^* (1)\epsilon (2)\epsilon (3)\sigma (4)_+>\simeq{1\over
|z_{41}|^{4/15}|z_{23}|^{8/5}}\ +\ {\displaystyle
C_{\sigma,\sigma^*,(25|25)}C_{\epsilon,\epsilon,(25|25)}\over
|z_{41}|^{-38/15}|z_{23}|^{-6/5}|z_{13}|^{28/5}}\ \ ,\eqno{(12a)}$$
and
$$\eqalign{<\sigma (1)&\sigma (2)\sigma^*(3)\sigma^*(4)_+>\simeq
{1+{\bf O}(z_{41},z_{23})\over |z_{41}|^{4/15}|z_{23}|^{4/15}}\ +\
 {\displaystyle |C_{\sigma^*,\sigma,\epsilon}|^2(1+{\bf O}(z_{41},z_{23}))\over
|z_{41}|^{-8/15}|z_{23}|^{-8/15}|z_{13}|^{8/5}}\cr
+&\left[\!\left[ {\displaystyle |C_{\sigma^*,\sigma,(25|25)}|^2\over
|z_{41}|^{-38/15}|z_{23}|^{-38/15}|z_{13}|^{28/5}}
+{\displaystyle |C_{\sigma^*,\sigma,(15|15)}|^2\over
|z_{41}|^{-86/15}|z_{23}|^{-86/15}|z_{13}|^{12}}\right. \right.\cr
&+{\displaystyle -|C_{\sigma^*,\sigma,(11|15)}|^2\over z_{41}^{2/15}
\bar{z}_{41}^{-43/15}z_{23}^{2/15}\bar z_{23}^{-43/15}\bar z_{13}^{6}}
+{\displaystyle -|C_{\sigma^*,\sigma,(15|11)}|^2\over \bar z_{41}^{2/15}
{z}_{41}^{-43/15}\bar z_{23}^{2/15} z_{23}^{-43/15} z_{13}^{6}}\cr
+&\left. \left. {\displaystyle -|C_{\sigma^*,\sigma,(25|21)}|^2\over
z_{41}^{-19/15}\bar z_{41}^{-4/15}z_{23}^{-19/15}\bar
z_{23}^{-4/15}z_{13}^{14/5}\bar z_{13}^{4/5}}
+ {\displaystyle -|C_{\sigma^*,\sigma,(21|25)}|^2\over  z_{41}^{-4/15}\bar
z_{41}^{-19/15}z_{23}^{-4/15}\bar z_{23}^{-19/15}z_{13}^{4/5}\bar
z_{13}^{14/5}}\right]\!\right]\ .}\eqno{(12b)}$$
In writing (11) and (12), we have used the transformations of the structure
constants under complex conjugation and cyclic permutations of indices (8). We
have also enclosed several terms of (12) in $\left[\!\left[...\right]\!\right]$
to indicate that corrections from other more divergent terms contribute with
the same powers of $z_{41}$ and $z_{23}$. The structure constants in the
$\left[\!\left[...\right]\!\right]$-terms are {\it more} simply calculated with
{\it other} 4-point functions.

These limit forms of the 4-point correlations will be compared with the
factorized form of correlation functions written in terms of conformal blocks.
This comparison gives the famous bootstrap equations that determine the
structure constants [3]. The last correlation of (12) hints that the $D_5$
correlations are not generally diagonal in t-channel conformal blocks unlike
the situation in the $A_5$ minimal model [4,12].

The operator product results of (11) and (12) were constrained by the $\bf Z_3$
symmetry of the 3-state Potts model. Thus, our solutions to the bootstrap
equations will respect this symmetry. In the $D$-series models, the doubling of
primary fields with identical conformal weights seems to destroy the uniqueness
of the solutions of these equations. The imposition of discrete symmetries
restores the uniqueness [6]. One must explicitly constrain the solutions by
imposing the discrete symmetries of the statistical model on the short distance
expansions   to obtain the correct critical correlation functions from
conformal field theory.

\vskip .2cm

\leftline{\bf 3. Conformal Blocks}

\vskip .2cm

The general technique for constructing the factorized 4-point correlations and
the relevant conformal blocks was described by Dotsenko and Fateev [12]. The
modifications necessary to apply this method to the nondiagonal $D_5$ model
were discussed by us [6]. We recall some details that aid in understanding how
results are extracted from these references.

The 4-point correlations can be written in a partially factorized form [3].
$$\langle \prod^4_{i=1}\Phi_{(r_i,s_i|r^\prime_i,s^\prime_i)}(z_i,\bar
z_i)_+\rangle =\sum_{l,k}A_{kl}F_k(z_1,...,z_4)\bar F_l(\bar z_1,...\bar z_4)\
. \eqno{(13)}$$
The conformal blocks, $ F_k(z_1,...,z_4)$ can be written through the Coulomb
gas formalism as integrals, over closed contours, of free bosonic correlations
[12].
$$\eqalign{F_k(z_1,...,z_4){\buildrel {R\rightarrow\infty}\over =}&\langle
\prod^3_{i=1}V_{a_{r_i,s_i}}(z_i)
V_{a_+ + a_- -a_{r_4,s_4}}(z_4)\prod^N_{k=1}\oint_{C(k)}V_{a_+}(v_k)\cr
&\times\prod^M_{l=1}\oint_
{D(l)}V_{a_-}(u_l)V_{-a_+-a_-}(R)\rangle R^{2(a_++a_-)^2}\cr} \eqno{(14)}$$
Here, the $V_{a_{r,s}}(z)$'s are bosonic vertex opertators and
$a_{r,s}={-1\over 2}([r-1]a_-+[s-1]a_+)$. They will describe blocks for primary
fields of the conformal theory having dimensions
$\Delta_{r,s}=a_{r,s}(a_{r,s}-a_+-a_-)$. The two special vectors $a_{\pm}$, for
which $\Delta_{a_\pm}\equiv 1$, define the screening operators of the Coulomb
gas representation. For $D_5$, we find that $a_+=\sqrt{5\over 6}$ and
$a_-=-\sqrt{6\over 5}$.
The explicit calculation of the conformal blocks of (14) simplifies when one
uses the $sl(2,{\bf C})$ invariance of the ground state. It implies that [6]:
$$F_k(z_1,...,z_4){\buildrel {\epsilon\rightarrow 0^+}\over =}\prod^4_{i=1}
\left({d\omega(z)\over dz}\right)^{\Delta_{r_i,s_i}}_{\displaystyle
|_{z=z_i}}F_k(0,z,1,{z_{34}
z_{41}\over z_{31}\epsilon})\ {\rm with} \ \omega (z)={z_{34}(z-z_1)\over
z_{31}(z-z_4 +\epsilon)}\eqno{(15)}$$

For the 3-state Potts model, the following Coulomb gas correspondences
exist.\footnote{$^4$}{In the following, we use the definition
$z\equiv\omega(z_2)={z_{12}z_{34}\over z_{13}z_{24}}$.}
For $\langle \sigma^*(1)\epsilon(2)\epsilon (3)\sigma (4)_+ \rangle$,
$a_{r_1,s_1}=a_{r_4,s_4}=-{1\over 2}a_--a_+$ and
$a_{r_2,s_2}=a_{r_3,s_3}=-{1\over 2}a_-$. For $\langle
\sigma(1)\sigma(2)\sigma^* (3)\sigma^* (4)_+ \rangle$,
$a_{r_1,s_1}= a_{r_4,s_4}= a_{r_2,s_2}=a_{r_3,s_3}=-{1\over 2}a_--a_+$.
Therefore, the Coulomb gas expression of (14)  with $N=0,\ M=1$, or one
$V_{a_-}$, gives the conformal blocks necessary to construct $\langle
\sigma^*(1)\epsilon(2)\epsilon (3)\sigma (4)_+ \rangle$, and the expression
with $N=2,\ M=1$ gives the blocks necessary for
$\langle\sigma(1)\sigma(2)\sigma^* (3)\sigma^* (4)_+ \rangle$.

{}From (15) and the free boson correlations, it is straightforward to rewrite
the blocks of (14) as closed contour integrals. The contour integral forms are
inconvenient, but they can be evaluated from simple integrals on the real line
[12]. The simple integrals are defined for unphysical exponents that guarantee
their convergence (see below). Analytic continuations in these exponents are
equal to the contour integral forms coming directly from (14). The simple
integral forms have one important property. They allow one to make short
distance expansions in the s- and t-channels that can be compared with the
operator product forms of (11) and (12). The analytic continuations necessary
to make these comparisons can be made directly on the short distance
expansions. The need to rely on analytic continuations makes this formalism
less useful for obtaining the 4-point correlations explicitly. Nevertheless, it
is very convenient for comparing with s- and t-channel short distance
expansions and thus sufficient to det

ermine structure constants.

With these remarks, we briefly describe the blocks necessary for our
calculations [6,12].
For $\langle \sigma^*(1)\epsilon(2)\epsilon (3)\sigma (4)_+ \rangle$, there are
two independent blocks having the form:\footnote{$^5$}{We introduce the blocks
${\tilde F}_q(z,a^\prime,b^\prime,c^\prime)$, without prefactors, for later
convenience when doing short distance expansions in the t-channel.}
$$\eqalign{F_{q}(z_1,...,z_4)=&{(z_{14})^{2/3}\over (z_{13}z_{24})^{4/5}}
z^{-2/5}(1-z)^{3/5}
\int du u^{a^\prime}|u-1|^{b^\prime} |u-z|^{c^\prime}\cr
\equiv& \left[{(z_{14})^{2/3}\over (z_{13}z_{24})^{4/5}}
z^{-2/5}(1-z)^{3/5}\right]{\tilde F}_q(z,a^\prime,b^\prime,c^\prime)\ \ \ .\cr}
\eqno{(16)}$$
The physical blocks correspond to the analytic continuation of the exponents of
this integral to the values $a^\prime=4/5$ and $b^\prime=c^\prime=-6/5$. Two
independent blocks are defined by the limits of integration. $F_1$ corresponds
to the block having the integral $\int^\infty_1du$ and $F_2$ to the block
having the integral $\int^z_0 du$.

The blocks for $\langle \sigma(1)\sigma(2)\sigma^* (3)\sigma^* (4)_+ \rangle$
come from the integral expressions:
$$\eqalign{F_{k,q}(z_1,...,z_4)=&{[z(1-z)]^{4/15}\over (z_{13}z_{24})^{2/15}}
\int dv_1\int dv_2\int du
\prod^{2}_{i=1}\left(|v_i|^{a} |v_i-1|^{b} |v_i-z|^{c}\right)\cr
&\times|u|^{a^\prime} |u-1|^{b^\prime} |u-z|^{c^\prime}
|v_1-v_2|^{g}\prod^{2}_{i=1}(v_i-u)^{-2}\cr
\equiv&\left[{[z(1-z)]^{4/15}\over (z_{13}z_{24})^{2/15}}\right]
\tilde{F}_{k,q}(z,a,b,c,g,a^\prime,b^\prime,c^\prime)\cr}\eqno{(17a)}$$
The physical blocks correspond to analytic continuation of the exponents to the
values $a^\prime=b^\prime=c^\prime=4/5$, $a=b=c=-2/3$ and $g=5/3$. The
expression of (17a) gives six independent blocks, $F_{k,q}$, with $k=1,2,3,$
and $q=1,2$. They are defined by the ranges of the integrals. For the values of
$q=1,2$, the ranges of the integration on "$u$" are defined to be
$\int^\infty_1 du$ and $\int^z_0 du$ respectively. For the values of $k=1,2,3$,
the ranges of the integral $\int dv_1\int dv_2$ are defined below.
$$\matrix{k=1\ \ \ \ \ \ & \int^\infty_1 dv_1\int^{v_1}_1 dv_2\cr
k=2 \ \ \ \ \ \ & \int^\infty_1 dv_1\int^z_0 dv_2\cr
k=3 \ \ \ \ \ \ & \int^z_0 dv_1 \int^{v_1}_0 dv_2\cr}\eqno{(17b)}$$

{}From (16) and (17), we can obtain expansions of the blocks in the s-channel,
$|z|\ll 1$, and in the t-channel, $|1-z|\ll 1$. They depend on the domains of
integration. For the blocks of (16), $F_k$, one gets the following expansions
in the s-channel.
$$\eqalign{{\tilde F}_1(z,a^\prime,b^\prime,c^\prime)=
N_1(a^\prime,b^\prime,c^\prime)(1+{\bf \rm O}(z))
=&B(-1-a^\prime-b^\prime-c^\prime,1+b^\prime)\cr
{\tilde F}_2(z,a^\prime,b^\prime,c^\prime)=
z^{1+a^\prime +c^\prime}N_2(a^\prime,b^\prime, c^\prime)(1+{\bf \rm O}(z))
=&z^{1+a^\prime +c^\prime}B(1+a^\prime,1+c^\prime)\cr
{\rm with}\ \ \ \ \ \ \ \ \ \ \ B(1+f,1+g)\equiv \int^1_0 du
u^f(1-u)^g=&{\Gamma (1+f) \Gamma (1+g)\over \Gamma (2+f+g)}}\eqno{(18)}$$
To compute the expansion in the t-channel, one needs the form of the monodromy
matrix $\bf [\alpha (1)]$. For the blocks of (16), it is defined by the
equation:
$$\tilde F_q(z,a^\prime,b^\prime,c^\prime)=\sum_{q^\prime}{\bf [\alpha(1)]}_{q
q^\prime}\tilde F_{q^\prime}(1-z,b^\prime,a^\prime,c^\prime)\eqno{(19a)}$$
The monodromy matrix $[{\bf \alpha(1)}]$ is easily calculated by rewriting the
conformal blocks as integrals over open contours [12]. One finds that:
$$[\bf \alpha(1)]=\left[\matrix{{\displaystyle sin(\pi a^\prime )\over
\displaystyle sin (\pi[b^\prime +c^\prime])} & {\displaystyle -sin(\pi
c^\prime)\over \displaystyle sin (\pi[b^\prime +c^\prime])}\cr
      {\displaystyle -sin(\pi [ a^\prime +b^\prime +c^\prime])\over
\displaystyle sin (\pi[b^\prime +c^\prime])}& {\displaystyle -sin(\pi
b^\prime)\over\displaystyle  sin (\pi[b^\prime +c^\prime])}\cr}\right]=
\left[\matrix{{\displaystyle -sin({\pi/5})\over\displaystyle  sin ({2\pi/ 5})}
& {\displaystyle sin({\pi/ 5})\over \displaystyle sin ({2\pi/ 5})}\cr
      +1 & {\displaystyle sin({\pi/5})\over\displaystyle
 sin ({2\pi/ 5})}\cr}\right]\eqno{(19b)}$$
Using $\bf [\alpha(1)]$ and expanding the right hand side of (19a) in powers of
$(1-z)$, one obtains the t-channel form of the blocks $F_q$.

Similarly, one easily obtains the expansion of the $F_{k,q}$'s, in the
s-channel, from the integral expressions of (17). Here, we only give the
results for the physical values, i.e. $a^\prime=b^\prime=c^\prime=4/5$,
$a=b=c=-2/3$ and $g=5/3$.\footnote{$^6$}{The short distance expansion of
$\tilde F_{k,q}(1-z,...)$, for $|1-z|\longrightarrow 0$, can be obtained from
the expansion of $\tilde F_{k,q}(z,...)$, for $|z|\longrightarrow 0$, by the
exchange $z_1\leftrightarrow z_3$, up to a phase which is independent of
$k,q$.}
$$\eqalign{ F_{1,1}(z_1,...,z_4)
=&{N_{1,1}z^{4/15}\over (z_{13}z_{24})^{2/15}}=
{{\cal J}_{1,2}(1/3,-2/3,5/6)\over (z_{12}z_{34})^{-4/15} z_{24}^{4/5}} \cr
F_{1,2}(z_1,...,z_4)
=&{N_{1,2}z^{43/15}\over (z_{13}z_{24})^{2/15}}={{\cal
J}_{0,2}(1/3,-2/3,5/6)B(9/5,9/5)\over (z_{12}z_{34})^{-43/15} z_{24}^{6}}\cr
F_{2,1}(z_1,...,z_4)
=&{N_{2,1}z^{-1/15}\over (z_{13}z_{24})^{2/15}}={{\cal J}_{1,1}(1/3,-2/3,5/6)
B(1/3,1/3)\over (z_{12}z_{34})^{1/15} z_{24}^{2/15}} \cr
 F_{2,2}(z_1,...,z_4)
=&{N_{2,2}z^{8/15}\over (z_{13}z_{24})^{2/15}}={{\cal J}_{1,1}(-2/3,-2/3,5/6)
B(4/3,1/3)\over (z_{12}z_{34})^{-8/15} z_{24}^{4/3}}\cr
F_{3,1}(z_1,...,z_4)
=&{N_{3,1}z^{19/15}\over (z_{13}z_{24})^{2/15}}={{\cal J}_{0,2}(-2/3,-2/3,5/6)
B(3/5,9/5)\over (z_{12}z_{34})^{-19/15} z_{24}^{14/5}} \cr
F_{3,2}(z_1,...,z_4)
=&{N_{3,2}z^{-2/15}\over (z_{13}z_{24})^{2/15}}={{\cal J}_{1,2}(-2/3,-2/3,5/6)
\over (z_{12}z_{34})^{2/15}}\cr } \eqno{(20)}$$
The direct calculation of the normalization constants $N_{k,q}$ is lengthy. By
 rewriting them in terms of ${\cal J}_{n,m}(\alpha,\beta,\rho)$'s, we can use
the results of Ref. [12] to determine them. The ${\cal
J}_{n,m}(\alpha,\beta,\rho)$'s are defined by the analytic continuation of the
following contour integral.
$$\eqalign{{\cal J}_{n,m}(\alpha,\beta,\rho)=&{1\over
n!m!}\prod^n_{i=1}\int^1_0
dt_it^{\alpha^\prime}(1-t_i)^{\beta^\prime}\prod^m_{j=1} \int^1_0 d\tau_j
\tau^\alpha_j (1-\tau_j)^\beta\cr
&\times\prod_{i<i^\prime}|t_i-t_{i^\prime}|^{2/\rho}
\prod_{j<j^\prime}|\tau_j-\tau_{j^\prime}|^{2\rho}
\prod^{n,m}_{i,j}(t_i-\tau_j +i0^+)^{-2}\cr}\eqno{(21)}$$
To rewrite the $N_{k,q}$'s in terms of ${\cal J}_{n,m}(\alpha,\beta,\rho)$ one
uses changes of variables like $v_j\equiv {1\over \tau_j}$ and $u\equiv {1\over
t_1}$ as well as the invariance of the integrand of (21) under $t_i
\longleftrightarrow t_{i^\prime}$ and
$\tau_j \longleftrightarrow \tau_{j^\prime}$. The calculation of ${\cal
J}_{n,m}(\alpha,\beta,\rho)$ makes use of the analytic properties of the
integrals under continuations of the exponents. The value of ${\cal J}_{n,m}$
is known when $\alpha=-\rho\alpha^\prime$ and $\beta=-\rho\beta^\prime$ [14].
These values are sufficient to fix the normalization constants coming from
conformal blocks. The six normalization constants of (20) have the following
form.
$$\matrix{N_{1,1}=5\sqrt{\pi\over 3} {\displaystyle
\Gamma(3/5)\Gamma(4/5)\left[\Gamma(1/6)\right]^2\over \displaystyle
\Gamma(2/5)\Gamma(5/6)} &
N_{1,2}={\displaystyle 5\sqrt{\pi}\over \displaystyle 3^{9/2}(13)}
{\displaystyle \left[\Gamma(1/6)\Gamma(4/5)\right]^2\over \displaystyle
\Gamma(5/6)\Gamma(3/5)}\cr
N_{2,1}=5{\displaystyle \Gamma(3/5)\Gamma(4/5)\left[\Gamma(1/3)\right]^4\over
\displaystyle \Gamma(2/5)\left[\Gamma(2/3)\right]^2}
& N_{2,2}={5\over 3}\left[{\displaystyle \Gamma(4/5)\over\displaystyle
\Gamma(2/3)}\right]^2{\displaystyle \left[\Gamma(1/3)\right]^4\over
\displaystyle \Gamma(3/5)}\cr
N_{3,1}={\displaystyle 5\sqrt{\pi}\over\displaystyle  7(3)^{3/2}}
{\displaystyle \Gamma(3/5)\Gamma(4/5)\left[\Gamma(1/6)\right]^2\over
\displaystyle \Gamma(2/5)\Gamma(5/6)}&
N_{3,2}=10\sqrt{\pi\over 3}{\displaystyle [\Gamma(4/5)\Gamma(1/6)]^2\over
\displaystyle \Gamma(5/6) \Gamma(3/5)}\cr}\eqno{(22)}$$

Finally, we must determine a second monodromy matrix ${[\bf \alpha(2,1)]}_{kq|
k^\prime q^\prime}$ to make the t-channel short distance expansion of the
blocks $\tilde F_{k,q}$. As in the previous case, the calculation of these
matrices involves rewriting (17a) as an integral on open contours in $v_i$ and
$u$ and then using the residue theorem to shift the contours (see Ref. [12]).
This allows one to express integrals like $\int^\infty_1dv_1\int^{v_1}_1dv_2$
in terms of integrals having the forms  $\int^{-\infty}_0dv_1\int^{v_1}_0 dv_2$
and $\int^z_1dv_1\int^{v_1}_1 dv_2$. These later integrals are equal to the
$\tilde F_{k,q}(1-z,b,a,c,g,b^\prime,a^\prime,c^\prime)$'s. One arrives at the
equation:
$$\tilde
F_{k,q}(z,a,b,c,g,a^\prime,b^\prime,c^\prime)=\sum_{k^\prime,q^\prime}{[\bf
\alpha(2,1)]}_{qk|k^\prime q^\prime}\tilde
F_{k^\prime,q^\prime}(1-z,b,a,c,g,b^\prime,a^\prime,c^\prime)\eqno{(23a)}$$

In applying these techniques to (17), an important simplification occurs if one
realizes that ${\bf[ \alpha(2,1)}]_{k q|
k^\prime q^\prime}={\bf[ \alpha(2)]}_{k k^\prime}\bigotimes {\bf
[\alpha(1)]}_{q q^\prime} $, i.e. the monodromy matrix factorizes. This
factorization is fairly easy to prove [14]. It follows from two facts. First,
the factors of $(v_i-u)$, in the integral forms for the conformal blocks lead
to double poles. Second, the fact that $a={-g\over 2} a^\prime, b={-g\over 2}
b^\prime$, and $c={-g\over 2}c^\prime$ allows one to write the remaining
integrals as total derivatives over closed contours. Thus, we can ignore the
$u$-dependence, in the integrand of (17a), while calculating ${\bf
[\alpha(2)]}_{k k^\prime}$, and we can ignore $v_i$-dependence while
calculating ${\bf [\alpha(1)]}_{q q\prime}$. The monodromy matrices ${\bf
[\alpha(1)]}_{q q\prime}$ and ${\bf [\alpha(2)]}_{k k^\prime}$ are found in the
first reference of [12]. The final form for the monodromy matrix at the
physical values of the exponents is:
$$\left[{\bf \alpha(2,1)}\right]_{k q | k^\prime q^\prime}
=\left[\matrix{ 1/2& -1 & 1/2 \cr -1/2 & 0 & 1/2 \cr
                                  1/2 & 1 & 1/2 \cr} \right]_{k k^\prime}
\bigotimes\left[\matrix{{\displaystyle -sin(\pi/ 5)\over\displaystyle  sin
(2\pi/ 5)} & {\displaystyle sin(\pi /5)\over \displaystyle sin (2\pi /5)}\cr
      +1 & {\displaystyle sin(\pi/5)\over\displaystyle
sin(2\pi/5)}\cr}\right]_{q q^\prime}\eqno{(23b)}$$
Now, we turn to the comparison of the conformal block form of the 4-point
correlations to the operator product expansions of Section Two.

\vskip .2cm

\leftline{\bf 4. Bootstrap Constraints and Stucture Constants}

\vskip .2cm

By comparing the factorized form of the 4-point functions, (13), to the
operator expansions in both the s- and t-channels, one can fix the constant
matrices $A_{kl}$. In the s-channel, this bootstrap procedure requires equating
the leading powers of $z_{12}$, $z_{34}$ of the conformal blocks (see (18) and
(20)) to the expansions in (11). In the t-channel, the procedure involves
comparing leading terms in $z_{41}$, $z_{23}$ (see (19) and (23)) to the
expansions in (12). The absolute normalization of all the structure constants
is fixed by the appearance of "1" in the t-channel expansions, (12), i.e. the
contribution of the 2-point function to the operator product expansions. It is
necessary to compare conformal block expressions to operator product expansions
in two channels in the $D$-series models, because 4-point correlations are
nondiagonal,  and the technique of requiring trivial monodromy invariance that
worked in the $A$-series is inapplicable [12].

Applying this procedure to the mixed correlation of (11a), one obtains the
following s-channel equation:
$$\left[\matrix{|C_{\sigma^*,\epsilon,\sigma}|^2 & 0 \cr  0 &
|C_{\sigma^*,\epsilon,\Omega}|^2 \cr}\right]=\left[\matrix{ N_1^2 A(1)_{11}  &
N_1N_2 A(1)_{12} \cr  N_1N_2 A(1)_{21} & N^2_2A(1)_{22}}\right]\eqno{(24a)}$$
Here, $N_q\equiv N_q(4/5,-6/5,-6/5)$.
These equations require that $A(1)_{12}=A(1)_{21}=0$. They also determine the
two structure constants in terms of $A(1)_{11}$ and $A(1)_{22}$. The t-channel
equations fix these remaining two components of $A(1)_{qq^\prime}$. One uses
(19) to transform (13) to a simple form in the t-channel and then expands in
powers of $z_{41}$ and $z_{23}$.  From (12a), (13), (18), and (19) one obtains
that:
$$\eqalign{\left[\matrix{C_{\sigma,\sigma^*,(25|25)}C_{\epsilon,
\epsilon,(25|25)} & 0 \cr 0 & 1\cr}\right]&=\left[\matrix{ N_1& 0 \cr 0 & N_2
\cr}\right][\alpha(1)]^T[A(1)][\alpha(1)]
\left[\matrix{ N_1& 0 \cr 0 & N_2  \cr}\right]=\cr
\left[\matrix{ N_1& 0 \cr 0 & N_2  \cr}\right]&\left[\matrix{ s^2A_{11} +A_{22}
&  -s^2A_{11} +sA_{22}\cr -s^2A_{11} +sA_{22} & s^2(A_{11} +A_{22})
\cr}\right]
\left[\matrix{ N_1& 0 \cr 0 & N_2  \cr}\right]}\eqno{(24b)}$$
Here, $N_q\equiv N_q(-6/5,4/5,-6/5)$, and $s\equiv{\displaystyle
sin(\pi/5)\over \displaystyle sin (2\pi/5)}$.
The diagonality of the t-channel equation requires that:
$$A(1)_{22}={sin(\pi/5)\over sin(2\pi/5)}A(1)_{11}\ .\eqno{(25a)}$$
Finally, the identity term on the right hand side of the t-channel equation
fixes the numerical value of $A(1)_{11}$.
$$1=N^2_{2}s^2[A(1)_{11}+A(1)_{22}]=A(1)_{11}B^2(-1/5,-1/5){\displaystyle
sin^2(\pi/5)\over \displaystyle sin^2(2\pi/5)}\left[1+{\displaystyle
sin(\pi/5)\over \displaystyle sin(2\pi/5)}\right]\ \ \ \ \eqno{(25b)}$$
Inserting (25) into the s-channel equation (24a), one obtains two nonzero
structure constants:
$$\matrix{|C_{\sigma^*,\epsilon,\sigma}|^2=&{\displaystyle sin^3(2\pi/5)\over
\displaystyle 4\ sin^2(\pi/5)[sin(2\pi/ 5) + sin(\pi/5)]}
{\displaystyle [\Gamma(3/5)]^4\over \displaystyle
[\Gamma(2/5)\Gamma(4/5)]^2}\cr |C_{\sigma^*,\epsilon,\Omega}|^2=&{\displaystyle
4\ sin^2(2\pi/5)\over \displaystyle 9\ sin(\pi/5)[sin(2\pi/5) +
sin(\pi/5)]}\cr}\eqno{(26)}$$
The signs of these two structure constants can be arbitrarily chosen to be
positive by fixing the signs of $\epsilon$ and $\Omega$.

Now, we repeat the same exercise for the $<\sigma (1)\sigma (2)\sigma^*
(3)\sigma^* (4)_+>$ correlation. Comparing the s-channel operator expansion,
(11b), with the conformal block form from (13) and (20), we see that only two
elements of $A(2,1)_{k q|k^\prime q^\prime}$ are nonzero, i.e.
$A(2,1)_{21|21}$ and $A(2,1)_{2 2|2 2}$. Thus, we can write the coefficient
matrix $A(2,1)_{k q|k^\prime q^\prime}$ in the form of a direct product.
$$A(2,1)_{k q|k^\prime q^\prime}=\left[\matrix{0&0&0\cr 0&1&0\cr 0&0&0
\cr}\right]_{k k^\prime}\bigotimes
\left[\matrix{\gamma&0\cr 0& \delta\cr}\right]_{q q^\prime}\eqno{(27)}$$
This also allows one to write that $\gamma\equiv
A(2,1)_{21|21}=N^{-2}_{2,1}|C_{\sigma,\sigma,\sigma}|^2$ and $\delta\equiv
A(2,1)_{22|22}=$  $N_{2,2}^{-2}|C_{\sigma,\sigma,\Omega}|^2$. The two nonzero
components of $A(2,1)_{k q|k^\prime q^\prime}$ are fixed by the t-channel
equation. Since $A(2,1)_{k q|k^\prime q^\prime}$ can be written in the same
direct product form as the monodromy matrix $[\alpha(2,1)]$ (see 23b), the
t-channel equations can also be written in the form of a direct product.
$$\eqalign{[\alpha(2,1)]&^T[A(2,1)][\alpha(2,1)]=\cr
&\left[\matrix{1/4&0&-1/4\cr0&0&0\cr-1/4&0&1/4\cr}\right]\bigotimes\left[
\matrix{ s^2A_{21|21}+A_{22|22}& -s^2A_{21|21}+sA_{22|22}\cr
-s^2A_{21|21}+sA_{22|22} & s^2(A_{21|21}+A_{22|22})\cr}\right]\cr}
\eqno{(28)}$$
Here, "$s$" has the same form as in (24b). This expression allows us to write
the t-channel form of $<\sigma(z_1)\sigma(z_2)\sigma^*(z_3)\sigma^*(z_4)_+>$ in
terms of conformal blocks:
$$\sum_{k,q,k^\prime,q^\prime}\left[{|1-z|^2|z|^2\over
|z_{13}||z_{24}|}\right]^{4/15}\bar{\tilde F}_{k,q}(1-\bar
z,...)\left([\alpha]^T[A][\alpha]\right)_{kq|k^\prime q^\prime}\tilde
F_{k^\prime,q^\prime}(1-z,...)\ \ . \eqno{(29)} $$
Comparing (12b) to this expression for $|1-z|,|z_{41}|,|z_{23}|\ll 1$, we find
that certain powers of $z_{41}$ and $z_{23}$ are absent. In making comparisons
between (12b) and (29), one must not neglect the possibility that corrections
to more divergent terms that are analytic in $z_{41}$ and $z_{23}$ can
contribute to nonleading divergences. Consistent with this limitation, one can
still show that the absence of primary fields like $\Phi_{(21|11)}$ and
$\Phi_{(11|21)}$ in the $D_5$ model implies that
$[\alpha^TA\alpha]_{11|32}=[\alpha^TA\alpha]_{32|11}=0$. Thus, the second
matrix of the direct product in (28) is diagonal, or equivalently
${\displaystyle A(2,1)_{21|21}sin(\pi/5)\over \displaystyle sin(2\pi/5)}
=A(2,1)_{22|22}$. The "1"-term, 2-point function contribution to (12b), again
determines $A(2,1)_{21|21}$. Using (20), (28), (29) and (12b), the resulting
equation can be written as:
$$N^2_{3,2} {\displaystyle sin^2(\pi/5)\over \displaystyle 4sin^2(2\pi/5)}
\left[{1+  {\displaystyle sin(\pi/5) \over \displaystyle
sin(2\pi/5)}}\right]A(2,1)_{21|21}=1\eqno{(30)}$$

Having fixed the complete form of the coefficient matrix $A(2,1)_{kq|k^\prime
q^\prime}$, we can read off $ C_{\sigma,\sigma,\sigma}$ and $
C_{\sigma,\sigma,\Omega}$ from the s-channel equations and
$C_{\sigma^*,\sigma,\epsilon} $ from the remaining t-channel equations. $$
\eqalign{C_{\sigma,\sigma,\sigma}&={2N_{2,1}\ sin^{3/2}(2\pi/5)\over  N_{3,2}\
sin(\pi/5)[sin(2\pi/5)+sin(\pi/5)]^{1/2}}\cr
C_{\sigma,\sigma,\Omega}&={2N_{2,2}\ sin(2\pi/5)\over  N_{3,2}\
(sin(\pi/5)[sin(2\pi/5)+sin(\pi/5)])^{1/2}}\cr
C_{\sigma^*,\sigma,\epsilon}&={N_{1,1}sin^{1/2}(2\pi/5)\over  N_{3,2}\
sin^{1/2}(\pi/5)}\cr}\eqno{(31)}$$
The other structure constants appearing in (12b) can only be found from the
t-channel equations by expanding the blocks of (17a) to nonleading orders in
$1-z$, a difficult task. They are more easily calculated from {\it other}
4-point correlations.
The explicit values of the three constants in (31) can be found with the aid of
(22). They are given below in Table 2. The signs of all the structure constants
of (26) and (31), except $C_{\sigma^*,\Omega,\epsilon}$, can be choosen to be
positive by fixing the phases of the three primary fields involved. The
remaining sign can be obtained by studying the correlation
$<\sigma(1)\sigma(2)\Omega(3)\epsilon(4)_+>$ (see Ref. [6]). Finally, the
equality of the expressions for $C_{\sigma^*,\sigma,\epsilon}$ from (26) and
(31) gives an interesting check of the method; they were obtained from
different correlation functions.

\vskip .2cm

\leftline{\bf 5. Conclusions}

\vskip .2cm

The EXACT values of the structure constants are given in Table 2.

\centerline{TABLE 2: Structure constants of 3-state Potts Model}
\vskip .1cm
\centerline{\vbox{\offinterlineskip\hrule
\halign{&\vrule#&\strut\quad\hfil#\hfil\quad\cr
&\omit&&\omit&\cr
&\omit&&\omit&\cr
height4pt&\omit&&\omit&\cr
height4pt&\omit&&\omit&\cr
&\omit&&\omit&\cr
&$ C_{\sigma,\sigma,\sigma}$&&${\displaystyle \sqrt{3}\ sin^{3/2}(2\pi/5)\over
\displaystyle \sqrt{\pi}\
sin(\pi/5)[sin(2\pi/5)+sin(\pi/5)]^{1/2}}\left[{\displaystyle
(\Gamma(3/5))^2(\Gamma(1/3))^4\Gamma (5/6)  \over \displaystyle
[\Gamma(2/3)\Gamma (1/6)]^2\Gamma(2/5)\Gamma(4/5)}\right]$&\cr
&\omit&&\omit&\cr
&\omit&&\omit&\cr
height4pt&\omit&&\omit&\cr
\noalign{\hrule}
height4pt&\omit&&\omit&\cr
&\omit&&\omit&\cr
&\omit&&\omit&\cr
& $C_{\sigma^*,\sigma,\epsilon}$&&  $ {\displaystyle sin^{3/2}(2\pi/5)\over
\displaystyle 2\ sin(\pi/5)[sin(2\pi/ 5) + sin(\pi/5)]^{1/2}}
{\displaystyle [\Gamma(3/5)]^2\over \displaystyle [\Gamma(2/5)\Gamma(4/5)]}$&
\cr
&\omit&&\omit&\cr
&\omit&&\omit&\cr
height4pt&\omit&&\omit&\cr
\noalign{\hrule}
height4pt&\omit&&\omit&\cr
&\omit&&\omit&\cr
&\omit&&\omit&\cr
&$C_{\epsilon,\epsilon,\epsilon}$&& 0&\cr
&\omit&&\omit&\cr
&\omit&&\omit&\cr
height4pt&\omit&&\omit&\cr
\noalign{\hrule}
height4pt&\omit&&\omit&\cr
&\omit&&\omit&\cr
&\omit&&\omit&\cr
& $ C_{\sigma,\sigma,\Omega}$&&${\displaystyle sin(2\pi/5)\over \displaystyle
\sqrt{3\pi}(sin(\pi/5)[sin(2\pi/5)+sin(\pi/5)])^{1/2}}\left[{\displaystyle
[\Gamma(1/3)]^4\Gamma(5/6)\over  \displaystyle
[\Gamma(2/3)\Gamma(1/6)]^2}\right]$&\cr
&\omit&&\omit&\cr
&\omit&&\omit&\cr
height4pt&\omit&&\omit&\cr
\noalign{\hrule}
height4pt&\omit&&\omit&\cr
&\omit&&\omit&\cr
& $C_{\sigma^*,\Omega,\epsilon}$&& ${\displaystyle\pm 2\ sin(2\pi/5)\over
\displaystyle 3\ [sin(\pi/5)(sin(2\pi/5) + sin(\pi/5))]^{1/2}} $   &\cr
height4pt&\omit&&\omit&\cr}
\hrule}}
The remaining structure constants of the critical 3-state Potts model can be
obtained by studing the bootstrap equations for other correlation
functions.\footnote{$^7$}{All the structure constants involving only primary
fields of the form $\Phi_{(1s|1s^\prime)}$ can be easily obtained from the
results of Ref. [6].} We have calculated some of the remaining constants and
have shown that they pass consistency requirements that eliminated some
possibilities for the discrete symmetry $\bf Z_2$ in $D_5$ conformal models
[6]. They are of less interest than the constants involving $\epsilon$ or
$\sigma$ that were calculated here. The $\epsilon$ or $\sigma$ fields are the
only naturally appearing fields in the Potts model. They couple to temperature
or "magnetic" field perturbations. The first two structure constants of Table 2
determine the behavior of the Potts model under such perturbations.

Finally, we mention that the Coulomb gas formalism had to be supplemented by
one physical imput, discrete symmetries, in comparing its predictions to
statistical models. These symmetries entered the calculations of the
correlations functions at the level of Fusion Rules. For the $D$-series
conformal models, they are necessary to make the solutions of the bootstrap
equations unique. All critical points of statistical models with complex order
parameters (and $c<1$) are described by $D$-series models. It appears very
likely that each $D$-series minimal conformal model will allow several
inequivalent sets of correlation functions that will correspond to statistical
models with different discrete symmetries. Finally, the measurement of these
universal quantities by experiments or simulations would be an interesting
confirmation of conformal field theories.

\vskip .2cm

J.M. wishes to thank the Laboratoire de Physique Th\'eorique de l'ENS de Lyon
for its hospitality during this work and acknowledges interesting discussions
with F. Delduc.

\vskip .2cm

\leftline{\bf 6. References}

\vskip .2cm

\item{[1]} R.J. Baxter, J. Phys. C: Solid State Phys. 6 (1973) L445; ibid, L94,
for a review see: R.J. Baxter, \underbar{Exactly Solved Models in Statistical
Mechanics}, (Academic Press 1989) 341-5.

\item{[2]} S. Alexander, Phys. Lett. A54 (1975) 353.

\item{[3]} A.A. Belavin, A.M. Polyakov, and A.B. Zamolodchikov, Nucl. Phys.
B241 (1984) 333.

\item{[4]} Vl.S. Dotsenko, Nucl. Phys. B235 (1984) 54; ibid, J. Stat. Phys. 34
(1984) 781.

\item{[5]} C. Itzykson and J.-B. Zuber, Nucl. Phys. B275 (1986) 580; A.
Cappelli, C. Itzykson, and J.-B. Zuber, Nucl. Phys. B280 (1987) 445; ibid,
Comm. Math. Phys. 113 (1987) 1; A. Kato, Mod. Phys. Lett. A2 (1987) 585; for a
review see: J. Cardy in \underbar{Fields, Strings and Critical Phenomena, Les
Houches 1988} (Eds.: E. Br\'ezin and J. Zinn-Justin).

\item{[6]} J. McCabe, T. Sami, and T. Wydro, "Structure Constants of the $D_5$
Chiral, Minimal Model," to appear in Int. J. of Mod. Phys. A.

\item{[7]} R.B. Potts, Proc. Camb. Phil. Soc. 48 (1952) 106, for a review see:
F.Y. Wu, Rev. of Mod. Phys. 54 (1982) 235.

\item{[8]} H.W.J. Bl\"ote, J.L. Cardy, and M.P. Nightingale, Phys. Rev. Lett.
56 (1986) 742; I. Affleck, Phys. Rev. Lett. 56 (1986) 746.

\item{[9]} L.V. Avdeev and B.-D. D\"orfel, J. Phys. A19 (1986) L13; F.
Woynarovich and H.-P. Eckle, J. Phys. A20 (1987) L97; C.J. Hamer, J. Phys. A19
(1986) 3335; C.J. Hamer, G.R.W. Quispel, and M.T. Batchelor, J. Phys. A20
(1987) 5677; C.J. Hamer and M.T.. Batchelor, J. Phys. A21 (1988) L173; F.
Woynarovich, Phys. Rev. Lett. 59 (1987) 259; for a review see: J. Cardy in
\underbar{Fields, Strings and Critical Phenomena, Les} \underbar{Houches 1988}
(Eds.: E. Br\'ezin and J. Zinn-Justin).

\item{[10]} D. Friedan, Z. Qiu, and S. Shenker, Phys. Rev. Lett. 52 (1984)
1575.

\item{[11]} J.L. Cardy and I. Peschel, Nucl. Phys. B300 (1988) 377.

\item{[12]} Vl.S. Dotsenko and V.A. Fateev, Nucl. Phys. B240 (1984) 312; ibid,
Nucl. Phys. B251 (1985) 691.

\item{[13]} A.M. Polyakov, JETP Lett. 12 (1970) 381.

\item{[14]} Vl.S. Dotsenko and V.A. Fateev, Nucl. Phys. B251 (1985) 691 (see
eqs. (A.1, A.2, A.35, A.36). We mention that the limits on the products in the
first form for ${\cal J}_{n,m}(\alpha,\beta,\rho)$ given in eq. (A.35) contain
some errors.)

\item{[15]} ibid at 706-7 (see also eqs. (A.5-7)).

\vfill\eject\end